# Origin of Immediate Damping of Coherent Oscillations in Photoinduced Charge Density Wave Transition

Yu-Xiang Gu[1,2,§], Wen-Hao Liu[1,2,§], Zhi Wang [1], Shu-Shen Li[1,2], Lin-Wang Wang[1]*, and Jun-Wei Luo[1,2]*

[1]*State Key Laboratory of Superlattices and Microstructures, Institute of Semiconductors, Chinese Academy of Sciences, Beijing 100083, China*

[2]*Center of Materials Science and Optoelectronics Engineering, University of Chinese Academy of Sciences, Beijing 100049, China*

[§]These authors contributed equally: Yu-Xiang Gu, Wen-Hao Liu.

*Email: lwwang@semi.ac.cn; jwluo@semi.ac.cn.

**Abstract:**

In stark contrast to the conventional charge density wave (CDW) materials, the one-dimensional CDW on the In/Si(111) surface exhibits immediate damping of the CDW oscillation during the photoinduced phase transition. Here, by successfully reproducing the experimentally observed photoinduced CDW transition on the In/Si(111) surface by performing real-time time-dependent density functional theory (rt-TDDFT) simulations, we demonstrate that photoexcitation promotes valence electrons from Si substrate to empty surface bands composed primarily of the covalent *p-p* bonding states of the long In-In bonds, generating interatomic forces to shorten the long bonds and in turn drives coherently the structural transition. We illustrate that after the structural transition, the component of these surface bands occurs a switch among different covalent In bonds, causing a rotation of the interatomic forces by about $\pi/6$ and thus quickly damping the oscillations in feature CDW modes. These findings provide a deeper understanding of photoinduced phase transitions.

Quasi-one dimensional indium atomic wires grown self-assembly on a silicon (111) surface, denoted as In/Si(111), have recently been extensively studied both experimentally and theoretically [1-14] since Yeom et al. [1] reported a reversible metal-to-insulator transition (MIT). At room temperature, the self-assembled In atomic wires on a Si(111) surface are composed of a pair of parallel zigzag In chains separated by a zigzag Si chain [2], resulting in a (4 × 1) unit cell, as shown in Fig. 1(a). This phase is in a metallic state as featured by three strongly dispersive, partially occupied, quasi-1D surface bands (denoted by $S_1$, $S_2$, and $S_3$ in Supplementary Fig. S3(a)). Below $T_c$ = 125 K, In atoms rearrange into distorted hexagons with an (8 × 2) reconstructed quadrupled unit cell (Fig. 1(b)) [3], accompanying the formation of 1D charge density wave (CDW) [1] owing to strong coupling between electron density modulation and periodic lattice distortion. In contrast to the metallic (4 × 1) phase, the symmetry reduced (8 × 2) phase opens a bandgap of 0.1-0.3 eV [1,4-7] at the $X_{8\times2}$ point and thus becomes an insulator. The structural transformation between the zigzag pattern and hexagon pattern is recognized experimentally after a long-time pursuit as a superposition of soft rotary phonon modes and shear phonon modes [8-14], which are also known as the CDW amplitude modes of the periodic lattice distortions [14]. Photoexcitation using ultrafast laser pulses has been recently demonstrated as an efficient route to manipulate MIT and to melt the 1D CDW in In/Si(111) on the timescale of lattice vibrations [13-21]. In striking contrast to all the other CDW materials [22-25], it exhibits rapid damping of the CDW mode oscillations during the photoinduced transition [15,21]. The underlying microscopic mechanism remains ambiguous, despite the postulates of an energy barrier hindering the immediate sliding back into the CDW mode [21] and strong damping of the CDW amplitudes through fast mode conversion caused by a rapid transfer of their energy to surface phonon modes of the Si substrate [15].

In this letter, we aim to reveal the microscopic mechanism underlying the swift suppression of the coherent CDW amplitude-mode oscillations by performing advanced real-time time-dependent density functional theory (rt-TDDFT) simulations [26,27]. Our rt-TDDFT calculations for the first time reproduce the experimentally observed dynamics of photoinduced CDW transition in In/Si(111) without *ad hoc* assumptions. We illustrate that the photoelectron population of the empty $S_1$ and $S_2$ bands generates atomic forces, driving the coherent motion of In atoms in CDW amplitude modes and resulting in the structural transition from the (8 × 2) phase to the (4 × 1) phase. We elucidate that the $S_1$ and $S_2$ bands are composed mainly of the bonding states of In bonds crossing two zigzag In chains in the (8 × 2) phase. However, in the (4 × 1) phase, the component of the $S_1$ and $S_2$ bands turns into the bonding states of In bonds within single zigzag In chains. We unravel that such bond switching prevents In atoms from motions in the CDW phonon modes via turning the direction of the atomic driving forces by about π/6.

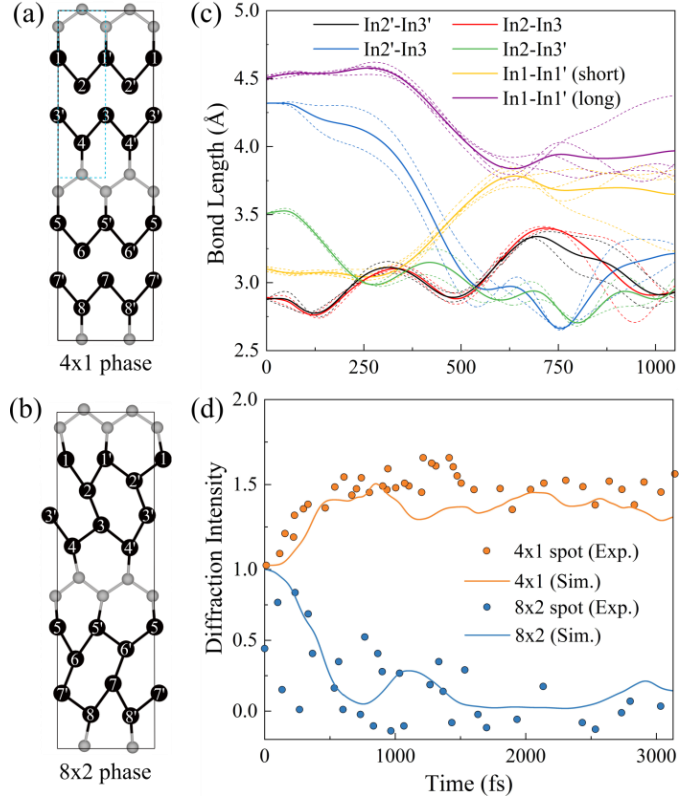

FIG. 1. The atomic configures of the In/Si(111) wire in (a) the zigzag (4×1) phase and (b) the hexagon (8×2) phase. The red dashed box in (a) indicates the (4×1) primary cell. The In atoms are numbered according to In lines from left to right and the prime superscript is used to distinguish two neighbors In atoms in each line. In the (4×1) phase, each zigzag In chain is composed of two In lines. (c) Dynamic evolution of the characteristic In-In bonds in the hexagon (8×2) phase following photoexcitation. (d) The time trajectory of the diffraction intensities of an (8×2) spot (indicative of the (8×2) structure) and a (4×1) spot (indicative of the (4×1) structure) obtained from our rt-TDDFT simulation in comparison with experimental data measured by ultrafast electron diffraction [29].

Figure 1 shows the rt-TDDFT simulation of structural dynamics of photoinduced phase transition in the In/Si(111) (8 × 2) surface under irradiation of 120 fs duration, 1.55 eV laser pulses with a fluence promoting about 3% of valence electrons from the valence bands to the conduction bands crossing the Fermi level $E_f$ (Supplementary Note1 and Fig. S1). We can see that it reproduces well the experimental observations [29]. The high-temperature zigzag (4 × 1) phase (Fig. 1(a)) has been recognized as connected to the low-temperature hexagon (8 × 2) phase (Fig. 1(b)) via the dimerizations of the outer In atoms in each zigzag In chain and of the inner In atoms across two zigzag In chains following the superposition of a shear distortion and two hexagon rotary distortions [9-14,28]. Figure 1(b) shows that the shear distortion between two zigzag In chains induces the dimerization of the inner In atoms as getting In2-In3, In2'-In3', In6-In7, and In6'-In7' bonds shorter (to 2.8 Å) and getting In2-In3', In2'-In3, In6-In7', and In6'-In7 bonds longer. The hexagon rotary distortion further breaks the identical bond length of In2-In3'/In2'-In3 and In6-In7'/In6'-In7 pairs into longer (4.3 Å) and shorter bonds (3.5 Å), and simultaneously, causes the dimerization of the outer In atoms by shortening the In1-In1' and In5-In5' bonds from 3.7 Å to 3.1 Å and elongating the In4-In4' and In8-In8' bonds from 3.7 Å to 4.5 Å. The atomic snapshots (shown in the Supplementary movie) exhibit that two (4 × 2) hexagon building blocks in the (8 × 2) phase undergo transitions simultaneously from hexagon structure to zigzag structure. For the sake of simplicity, we will examine one of two (4 × 2) hexagon blocks hereafter (unless stated otherwise).

Figure 1(c) shows that the outer In dimers start dissolution at $t = 300$ fs following the photoexcitation, manifesting as their bond lengths ($d_{1-1'} = d_{4-4'} = 3.1$ Å) increasing along with the shortening of their long counterparts. At $t = 600$ fs, both dimers and their long counterparts have the same bond length (3.7 Å) as their equilibrium bond length in the zigzag (4 × 1) phase. Whereas, the dissociation of the inner dimers (In2-In3 and In2'-In3') begins (earlier) at 125 fs after experiencing a shortening immediately following the photoexcitation. Meanwhile, their long counterparts (In2'-In3 and In2-In3' bonds) come into being shorter at 100 fs. Although the In2'-In3 bond ($d_{2'-3} = 4.3$ Å) is much longer than the In2-In3' bond ($d_{2-3'} = 3.5$ Å), they encounter a crossing in bond length at around 540 fs, approaching the same length as their equilibrium bond length in the zigzag (4 × 1) structure. After getting crossed, a slight overshoot takes place due to the continued elongating of the inner dimers and shortening of their long counterparts until 750 fs. They then rebound slightly (but never fully back) toward the (8 × 2) structure. We can deduce that the (8 × 2) - (4 × 1) transition occurs at around $t = 600$ fs despite the inner and outer In dimers reaching their exact equilibrium positions in the zigzag (4 × 1) phase at slightly different times ($t = 540$ and 600 fs, respectively). This deduction can be further confirmed by inspecting the snapshots of atomic configurations at $t = 60, 220, 300, 400, 500$, and 600 fs (shown in Fig. 3(a), a higher temporal resolution is also given in the Supplementary movie accompanied by corresponding band structure). One can see that the atomic configuration at $t = 600$ fs is truly close to the zigzag (4 × 1) structure, as shown in Fig. 1(a), with deviation invisible to the naked eye.

To make a direct comparison with the experimental electron diffraction data, we have to obtain the time-dependent diffraction intensity based on the atomic trajectories produced by the rt-TDDFT simulations according to the Debye-Waller factor [30,31],

$$I(t) = \exp[-Q^2 <u^2(t)>/3]. \tag{2}$$

Where $Q$ is the magnitude of the reciprocal lattice vector for the diffraction spot, and $u^2(t)$ is the squared displacement of atoms from their ideal lattice positions at time $t$ after photoexcitation (Supplementary Fig. S2). Figure 1(d) exhibits that the simulated diffraction intensities at both the (8×2) and (4×1) spots are in great agreement with the experimental data over the whole investigated period from 0 to 3000 fs. Specifically, the simulated diffraction intensity at the (8×2) spot drops sharply within 600 fs and at 750 fs reaches the minimum, which indicates the complete quenching of the hexagon (8×2) structure. Meanwhile, the diffraction intensity at the (4×1) spot grows up fast within 500 fs and then gradually increases to the maximum at 1000 fs. The slow increase of the (4×1) spot indicates the system in the vicinity of the zigzag (4×1) structure. Hence, we have demonstrated that the (8×2)-(4×1) structural transition deduced from the diffraction pattern is consistent with that obtained from the bond dynamics shown in Fig. 1(c).

Our rt-TDDFT simulation predicted that the photoinduced structural transition is completed within 600 fs, in excellent agreement with experimentally measured timescales of ~700 fs [15] and 660 ± 120 fs [13] from the lattice dynamics using time-resolved reflection high-energy electron diffraction (tr-RHEED) and of ~ 660 fs [13] and ~ 700 fs [18] from the band structure dynamics using time- and angle-resolved photoemission spectroscopy (tr-ARPES). From the band structure evolution shown in the Supplementary movie, we further find that the bandgap closure of the insulating (8×2) phase occurred at about 200 fs earlier than the completion of the structural transition, which is also consistent with the tr-ARPES observations [13,18]. Overall, to the best of our knowledge, we for the first time have theoretically reproduced the experimental observations of the photoinduced phase transition of In/Si(111) wires from the insulating (8×2) phase to the metallic (4×1) phase without *ad hoc* assumptions.

Particularly, our rt-TDDFT simulations validate all experimental findings [13,15,19,21] that the diffraction intensity stabilizes without any indications of coherent oscillatory behavior after the melting of the 1D CDW phase, which is in sharp contrast to the conventional photoinduced CDW transitions [15,25,32-36]. According to the Peierls picture, the photoinduced phase transitions are due to the photoexcitation of the displacive soft phonon modes that are connected to the lattice Peierls distortion. Because the damping of a phonon will take place on a much longer timescale than the phonon's period, these soft phonons will drive back-and-forth transitions between two transitional phases, generating coherent oscillation. In quasi-one-dimensional CDW $K_{0.3}MoO_3$ [32,33] and $Pr_{0.5}Ca_{0.5}MnO_3$ [34] as well as two-dimensional CDW 1T-TiSe$_2$ [25], 1T-TaS$_2$ [35], and 1T-TaSe$_2$ [36], the coherent oscillations in diffracted X-ray intensity controlled by photoexcitation have already been seen and discussed. In the In/Si(111) surface, the shear distortion (0.55~0.66 THz) and hexagon rotary distortion (0.82~0.84 THz) [9,11,12,14] are recognized as two displacive soft phonon modes (also termed as CDW amplitude modes in literature) that connect the hexagon (8×2) and zigzag (4×1) structures (phases) [13,14]. Therefore, the quick damping of the coherent oscillations is unexpected in the CDW phase transitions [25,32-36].

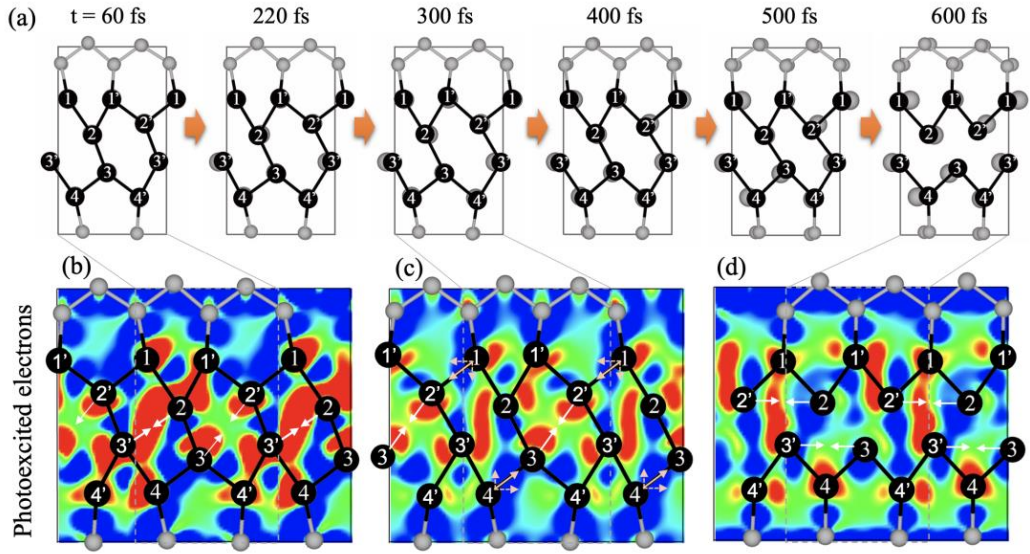

FIG. 2. Time evolution of atomic structure and photoexcited electrons of In/Si(111) in the (8 × 2) phase following photoexcitation. (a) Snapshots of atomic positions at 60, 220, 300, 400, 500, and 600 fs after photoexcitation. Big black dots represent In atoms (whose initial positions before photoexcitation are marked by grey color to guide the eye) and small grey dots for Si atoms. (b-d) The distributions of the photoexcited electrons at $t = 60$, 300, and 600 fs, respectively, following photoexcitation. The white arrows denote the photoexcitation-generated interatomic forces via the photoelectron population of surface conduction bands. Pink solid arrows in (c) indicate the forces pulling the In1 and In4 atoms resulting indirectly from the shrinkage of the In2'-In3 bond. Their decomposition components (vertical and horizontal directions) are represented by pink dashed arrows.

To gain deeper insight into the nature of the photoinduced transition, it is better to uncover the photoinduced atomic forces driving atoms in collective and directed motions that give rise to the CDW amplitude phonon modes. Fig. 2 (b) shows the real-space distribution of photoexcited electrons (holes given in Supplementary Fig. S4 and Fig. S5) that indicates that the 1.55 eV femtosecond laser pulses promote electrons from the valence bands of Si substrate to the empty surface bands of the In/Si(111) (8 × 2) structure. Considering only a small portion of photoholes located in the regime of In wires, we expect photoelectrons rather than photoholes to be the primary factor driving the phase transition. This expectation is opposite to that of the earlier AIMD simulations [15,18] in which photoexcited electrons and holes are artificially assumed to be in equilibrium distribution with a high electronic temperature. To prove it, we inspect the correlation between photoelectrons and structural evolution. Figure 2(b) displays that at 60 fs the maximum occupancy of photoelectrons is on the In2-In3' bond with a tiny component on the In2'-In3 bond. Figure 2(a) indeed shows that among all In atoms, only In2 and In3' (grey dots for their initial positions) undergo visible displacements in the period from 60 to 300 fs. We have also elucidated that, in the (8 × 2) phase, the (empty) conduction $S_2$ and $S_1$ bands mainly consist of the In $5p$ bonding states of the In2-In3' bond and the In2'-In3 bond, respectively (Fig. 3 and Supplementary Fig. S3). The large component of initially photoexcited electrons occupies the high-lying $S_2$ band, yielding photoelectrons that have a peak on the In2-In3' bond as shown in Fig. 2(b). It generates an attractive force in the In2-In3' and In2'-In3 bonds since the excited system will gain energy from lowering the energy of occupied bonding states via shortening the bond length [30,37,38]. Since the photoinduced interatomic force has a magnitude proportional to the carrier occupancy number, the In2-In3' bond has a larger interatomic force than the In2'-In3 bond due to its higher photoelectron density [30,31,39]. As expected, Figure 2(a) shows that the attractive force pulls In2 and In3' atoms closer in a much larger magnitude than that between In2' and In3 atoms within 300 fs.

As the $S_2$ band goes down in energy along with the shortening of the In2-In3' bond, the hot photoelectrons will relax to the low-lying $S_1$ band. As a consequence, Fig. 2(c) shows that the high photoelectron occupation changes from In2'-In3 bond to In2'-In3 bond at 300 fs. Such charge transfer remarkably enhances the attractive force in the In2'-In3 bond, which pulls the In2' and In3 atoms to move relative to one another along the bond (shown in Fig. 2(a)) and thus shortening the In2'-In3 bond (Fig. 1(c)) in the period from 300 to 500 fs. Whereas, in this period, the In2 and In3' atoms have no further movements due to the photoinduced attractive force being compensated by enhanced electrostatic repulsion of electron clouds. At 600 fs, the system achieves the zigzag (4 × 1) phase (Fig. 2(d)) following the de-dimerization of the outer In dimers along with the direction switch of boundary In-Si bonds perpendicular to the In wire (detailed in the Supplementary movie). Note that, during the (8 × 2) to (4 × 1) structural transition, In1 (In4) exhibits a leftward (rightward) displacement leaving behind its dimer's counterpart In1' (In4') having almost invisible displacement from its initial position. It implies that the atomic forces driving the movements of In1 and In4 arise indirectly from the shrinkage of the In2'-In3 bond, as indicated in Fig. 2(c). Their vertical components of atomic driving forces are compensated by Si-In bonds, leading to no vertical displacement.

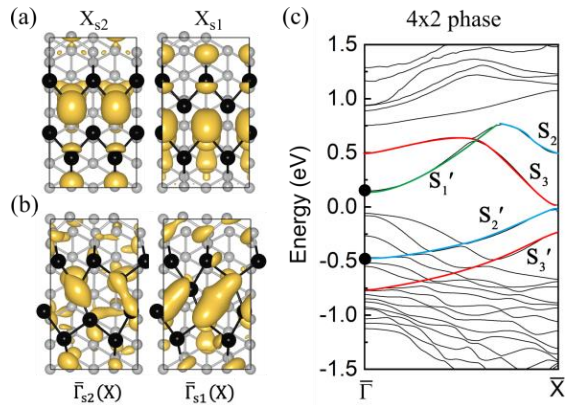

FIG. 3. The wavefunctions of the $S_1$ and $S_2$ bands at $X_{4x1}$ point in the (4×1) phase (b) and in the (4×2) phase (c) in which they are folded to the $\bar{\Gamma}$ point, accompanying the top view of the corresponding atomic structure. Here, the black circles represent In atoms and the gray circles represent Si atoms. (c) First-principles band structure of the In/Si(111) (4×2) structure (half of the (8×2) structure).

Interestingly, Figure 3 shows a character switch of the $S_1$ and $S_2$ bands from the In 5$p$ bonding states of In2-In3' and In2'-In3 dimers in the (8×2) phase to that of In2-In2' and In3-In3' bonds in the (4×1) phase (also see Supplementary Note3 for details). Because in the zigzag (4 × 1) phase both In2-In2' and In3-In3' bonds are parallel to the In wire, one can see that the photoelectron pattern undergoes a corresponding switch once the system achieves the (4 × 1) phase at $t$ = 600 fs, as shown in Fig. 2(b)-(d). As expected, Figure 2(d) shows that the directions of photoinduced atomic forces switch to parallel to the In wire by a rotation of about $\pi/6$. We can conclude that a bond switch in surface bands explains the disappearance of atomic forces driving the rebound back to the (8 × 2) phase, resulting in no oscillations of the CDW transition modes [13,15,21]. This microscopic is somewhat related to the argument of an energy barrier hindering the immediate sliding back into the CDW mode [21] but different from the argument of fast mode conversion caused by a rapid transfer of their energy to surface phonon modes of the Si substrate [15].

In summary, we have successfully reproduced the experimental observations of photoinduced phase transition in the In/Si(111) (8×2) structure by performing advanced rt-TDDFT simulations. We find that the photoelectron population of the empty surface $S_1$ and $S_2$ bands generates atomic driving forces for the coherent motion of In atoms in CDW amplitude modes, resulting in the transition from the (8 × 2) phase to the (4 × 1) phase. A bond switch in the surface $S_1$ and $S_2$ bands causes rotation of the atomic driving forces by about $\pi/6$, leading to the immediate damping of the CDW amplitude phonon modes.

**ACKNOWLEDGMENTS**

The work was supported by the National Natural Science Foundation of China (NSFC) under Grant Nos. 11925407 and 61927901, the Key Research Program of Frontier Sciences, CAS under Grant No. ZDBS-LY-JSC019, CAS Project for Young Scientists in Basic Research under Grant No. YSBR-026, the Strategic Priority Research Program of the Chinese Academy of Sciences under Grant No. XDB43020000, and the key research program of the Chinese Academy of Sciences under Grant No. ZDBS-SSW-WHC002.